\documentclass[12pt]{article}
\usepackage{epsf}
\title{ {\bf
The effect of antisymmetric tensor unparticle mediation on the
charged lepton electric dipole moment}}
\author{\vspace{1cm}\\
        {\bf E. O. Iltan}
        \thanks{E-mail address:
        eiltan@newton.physics.metu.edu.tr}
 \\
        Physics Department, Middle East Technical University \\
        Ankara, Turkey\\}

\date{}

\begin{document}
\setlength{\baselineskip}{24pt}
\maketitle
\setlength{\baselineskip}{7mm}
\begin{abstract}
We study the contribution of antisymmetric tensor unparticle
mediation to the charged lepton electric dipole moments and
restrict the free parameters of the model by using the
experimental upper bounds. We observe that the charged lepton
electric dipole moments are strongly sensitive to the the scaling
dimension $d_U$ and the fundamental scales $M_U$ and $\Lambda_U$.
The experimental current limits of electric dipole moments are
reached for the small values of the scaling dimension $d_U$.
\end{abstract}
\thispagestyle{empty}
\newpage
\setcounter{page}{1}
%
The CP violation which leads to the unequal amounts of matter and
antimatter in the universe needs more accurate theoretical
explanation. The electric dipole moments (EDMs) of fermions are
driven by the CP violating interaction and, therefore, their
search, especially the charged lepton EDMs\footnote{They are clean
theoretically since they are free from strong interactions.}, is
worthwhile in order to understand the CP violation mechanism. The
current experimental limits of the electron, muon and tau EDMs are
$d_e =(0.7\pm 0.7)\times 10^{-27} e\, cm$ \cite{Commins}
$d_{\mu} =(3.7\pm 3.4)\times 10^{-19} e\, cm$ \cite{Bailey} and
Re[$d_{\tau}$]$= -0.22$ to $0.45 \times   10^{-16} e\, cm$;
Im[$d_{\tau}$]$= - 0.25$ to $0.008  \times   10^{-16} e\, cm$
\cite{KInami}, respectively.
These experimental results stimulate the search of the lepton EDMs
in the framework of various theoretical models. In the standard
model (SM) the source of  the CP violation and, therefore the EDM,
is the complex Cabibo Kobayashi Maskawa (CKM) matrix in the quark
sector and the lepton mixing matrix in the lepton sector. However
the EDM predictions in the SM are negligible and far from their
current experimental limits. Therefore one goes beyond the SM such
as multi Higgs doublet models (MHDM), supersymmetric model (SUSY)
\cite{Schmidt}, left-right symmetric model, the seesaw model, the
models including the extra dimensions and noncommutative
effects,... etc., in order to get the additional CP violating
phase (see for example \cite{Iltmuegam}-\cite{IltanNonCom}).
Another possibility for a new CP violating phase is to consider
the recent unparticle idea which is proposed by Georgi
\cite{Georgi1, Georgi2}. Unparticles are new degrees of freedom
arising from the SM-ultraviolet sector interaction at some scale
$M_U$ and, because of the scale invariance, they are massless and
have non integral scaling dimension $d_U$, around the scale
$\Lambda_U\sim 1.0\,TeV$. The effective interaction of the
SM-ultraviolet (UV) sector at the scale $M_U$ reads
\begin{equation}
{\cal{L}}_{eff}= \frac{C_n}{M_U^{d_{UV}+n-4}}\,O_{SM}\,O_{UV}\,,
\label{eff1}
\end{equation}
with the scaling dimension $d_{UV}$ of the UV operator
\cite{BankZaks} and, around the scale $\Lambda_U$, it appears as
(see \cite{Rajaraman}, \cite{Tae} and references therein)
\begin{equation}
{\cal{L}}_{eff}=
\frac{C^i_n}{\Lambda_n^{d_{U}+n-4}}\,O_{SM,i}\,O_{U}\,,
\label{eff2}
\end{equation}
where
\begin{equation}
\Lambda_n=\Bigg(\frac{M_U^{d_{UV}+n-4}}{\Lambda_U^{d_{UV}-d_U}}\Bigg)^{
\frac{1}{d_U+n-4}} \,, \label{eff3}
\end{equation}
and $n$ is the scaling dimension of SM operator of type $i$. Here
the scale $\Lambda_n$ is sensitive to the scaling dimension $n$ of
the SM operator $O_{SM,i}$ \cite{Rajaraman, Tae} and depends on
the fundamental scales $M_U$, $\Lambda_U$\footnote{$\Lambda_2 <
M_U < \Lambda_4 < \Lambda_3$ with the choice $1< d_U< 2< d_{UV}$
(see \cite{Rajaraman}).}.

In the present work, we consider that the new CP violating phase
is coming from the effective unparticle fermion interaction and we
predict the charged lepton EDMs (see \cite{IltanUnp} for the
scalar unparticle contribution to the charged lepton EDM). Here we
assume that the antisymmetric tensor unparticle mediation gives
the contribution to the lepton EDM\footnote{The contribution of
the antisymmetric tensor unparticle mediation to the muon
anomalous magnetic dipole moment and its effects in $Z$ invisible
decays and the electroweak precision observable $S$ has been
predicted in \cite{Tae}.} by respecting the following conditions:
\begin{itemize}
\item  The scale $\Lambda_n$ in the effective Lagrangian depends
on the dimension of the SM operator $O_{SM,i}$,
\item antisymmetric tensor unparticle-lepton couplings are
complex,
\item  the scale invariance is broken at some scale $\mu$ after
the electroweak symmetry breaking due to the additional
interaction $\sim
\frac{\lambda_2}{\Lambda_2^{du-2}}\,O_S\,H^\dagger\,H$ where $H$
($O_S$) is the SM Higgs (scalar unparticle operator which exists
with the antisymmetric tensor unparticle) \cite{PJFox, Kikuchi}.
\end{itemize}
The two point function of antisymmetric tensor unparticle reads
(see Appendix for details)
\begin{eqnarray}
\int\,d^4x\,
e^{ipx}\,<0|T\Big(O^{\mu\nu}_U(x)\,O^{\alpha\beta}_U(0)\Big)0>=
i\,\frac{A_{d_U}}
{2\,sin\,(d_U\pi)}\,\Pi^{\mu\nu\alpha\beta}(-p^2-i\epsilon)^{d_U-2}
\, , \label{propagator}
\end{eqnarray}
where the factor $A_{d_U}$ is
\begin{eqnarray}
A_{d_U}=\frac{16\,\pi^{5/2}}{(2\,\pi)^{2\,d_U}}\,
\frac{\Gamma(d_U+\frac{1}{2})} {\Gamma(d_U-1)\,\Gamma(2\,d_U)} \,
. \label{Adu}
\end{eqnarray}
Here $\Pi^{\mu\nu\alpha\beta}$ is the projection operator
\begin{eqnarray}
\Pi_{\mu\nu\alpha\beta}=
\frac{1}{2}(g_{\mu\alpha}\,g_{\nu\beta}-g_{\nu\alpha}\,g_{\mu\beta})
\, , \label{projection}
\end{eqnarray}
and it can be divided into the transverse and the longitudinal
parts as
\begin{eqnarray}
\Pi^T_{\mu\nu\alpha\beta}=
\frac{1}{2}(P^T_{\mu\alpha}\,P^T_{\nu\beta}-P^T_{\nu\alpha}\,P^T_{\mu\beta})\,
, \,\,\,\,\,\,
\Pi^L_{\mu\nu\alpha\beta}=\Pi_{\mu\nu\alpha\beta}-\Pi^T_{\mu\nu\alpha\beta}
\, , \label{projectionTL}
\end{eqnarray}
with $P^T_{\mu\nu}=g_{\mu\nu}-p_{\mu}\,p_{\nu}/{p^2}$ (see for
example \cite{Tae} and references therein). Furthermore, the scale
invariance breaking at the scale $\mu$ results in that the
antisymmetric tensor unparticle propagator is modified. The
propagator is model dependent (see for example \cite{ADelgado} for
the scalar unparticle case) and we consider the one in the simple
model \cite{PJFox, ARajaraman}:
\begin{eqnarray}
\int\,d^4x\,
e^{ipx}\,<0|T\Big(O^{\mu\nu}_U(x)\,O^{\alpha\beta}_U(0)\Big)0>=
i\,\frac{A_{d_U}}
{2\,sin\,(d_U\pi)}\,\Pi^{\mu\nu\alpha\beta}(-(p^2-\mu^2)-i\epsilon)^{d_U-2}
\, . \label{propagatormu}
\end{eqnarray}
Here $\mu$ is the scale where unparticle sector changes in to the
particle sector.

Now we start with the effective Lagrangian responsible for the EDM
of charged leptons\footnote{Here we used the effective Lagrangian
given in \cite{Tae} and choose the unparticle-lepton coupling
complex in order to switch on the CP violation. In this equation
$H$ is the Higgs doublet, $g$ and $g'$ are weak couplings,
$\lambda_B$ and $\lambda_W$ are the unparticle-field tensor
couplings, $B_{\mu\nu}$ is the field strength tensor of the
$U(1)_Y$ gauge boson $B_\mu=c_W\,A_\mu+s_W\,Z_\mu$ and
$W^a_{\mu\nu}$, $a=1,2,3$, are the field strength tensors of the
$SU(2)_L$ gauge bosons with $W^3_\mu=s_W\,A_\mu-c_W\,Z_\mu$ where
$A_\mu$ and $Z_\mu$ are photon and Z boson fields respectively. }:
\begin{eqnarray}
{\cal{L}}_{eff}&=&\frac{g'\,\lambda_B}{\Lambda_2^{d_U-2}}\,B_{\mu\nu}\,
O^{\mu\nu}_U+\frac{g\,\lambda_W}{\Lambda_4^{d_U}}\,(H^\dagger\,\tau_a
\,H)\, W^a_{\mu\nu}\, O^{\mu\nu}_U  \nonumber \\
&+& \frac{y_l}{\Lambda_4^{d_U}}\Big(\lambda_l\, \bar{l}_L\,H
\,\sigma_{\mu\nu}\,l_R+\lambda_l^*\,\bar{l}_R\,H^\dagger
\,\sigma_{\mu\nu}\,l_L \Big)\, O^{\mu\nu}_U\, ,
\label{lagrangiantensor}
\end{eqnarray}
with the lepton field $l$ and the complex coupling
$\lambda_l=|\lambda_l|\,e^{i\,\theta_l}$ where $\theta_l$ is the
CP violating parameter.

\begin{figure}[htb]
\vskip 5.0truein \centering \epsfxsize=2.8in
\leavevmode\epsffile{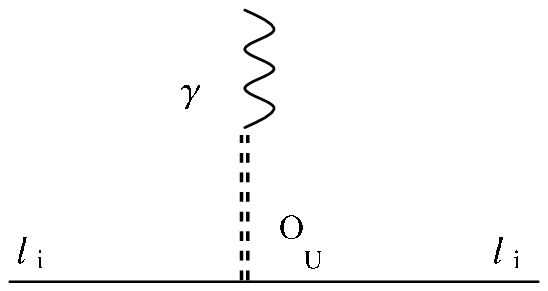} \vskip -1.0truein \caption[]{Tree
level diagram contributing to the EDM of charged lepton due to
tensor unparticle. Wavy (solid) line represents the
electromagnetic field (lepton field) and double dashed line the
tensor unparticle field.} \label{fig1}
\end{figure}

The effective EDM interaction for a charged lepton $l$ reads
\begin{eqnarray}
{\cal L}_{EDM}=i d_l \,\bar{l}\,\gamma_5 \,\sigma^{\mu\nu}\,l\,
F_{\mu\nu} \,\, , \label{EDM1}
\end{eqnarray}
where $F_{\mu\nu}$ is the electromagnetic field tensor and
'$d_{l}$', which is a real number by hermiticity, is the EDM of
the charged lepton. Finally, the effective Lagrangian in
eq.(\ref{lagrangiantensor}) leads to the EDM of charged leptons
$l$ after electroweak breaking as (see Appendix for details):
\begin{eqnarray}
d_l= -i(\lambda_{l}-\lambda^*_{l})\,
\frac{e\,\mu^{2\,(d_U-2)}\,\,A_{d_U}\,m_l\,}{2\,\,sin\,(d_U\pi )\,
\Lambda_4^{d_U}}\,\Bigg(\frac{\lambda_B}{\Lambda_2^{d_U-2}}-
\frac{v^2\,\lambda_W}{4\,\Lambda_4^{d_U}} \Bigg)\, ,
\label{EDMtensor}
\end{eqnarray}
where $v$ is the vacuum expectation value of the SM Higgs $H^0$.
\\ \\
{\Large \textbf{Discussion}}
\\ \\
In this section we predict the intermediate antisymmetric tensor
unparticle contribution (see Fig.\ref{fig1}) to the charged lepton
EDMs by considering that the CP violating phase is carried by the
tensor unparticle-charged lepton couplings and try to restrict the
free parameters of the model by using the experimental upper
bounds of the charged lepton EDMs. The scaling dimension of UV
operator $O_{UV}$ (the unparticle operator $O_U$) $d_{UV}$
($d_U$), the fundamental scales of the model, namely the
interaction scale $M_U$ of the SM-ultraviolet sector  and
interaction scale $\Lambda_U$ of the SM-unparticle sector  and the
scale $\mu$ which is responsible for the flow of unparticle sector
in to the particle one are among the free parameters. In our
numerical calculations  we choose the scale dimension $d_U$ in the
range\footnote{For antisymmetric tensor unparticle the scale
dimension should satisfy $d_U>2$ not to violate the unitarity (see
\cite{Grinstein}). Here we assumed that the scale invariance is
broken at some scale $\mu$ and the restriction on the values of
$d_U$ is more relaxed. We used the simple model \cite{PJFox,
ARajaraman} to define the new propagator. Since this model ensures
a connection with the particle sector, we choose $d_U$ in the
range $1< d_U <2$ and when $d_U$ tends to one one reaches the
particle sector and the connection is established. Since this
choice brings a rough connection between two sectors, unparticle
and particle sectors, we believe that it is worthwhile to study
even if it needs more careful analysis whether its is consistent
with the QFT.} $1< d_U <2$ and $d_{UV}>d_U=3$ (see
\cite{Rajaraman} and \cite{Tae}) and we choose $\mu\sim 1.0\,GeV$.
The couplings $\lambda_B$, $\lambda_W$ and $\lambda_l$ are other
free parameters which should be restricted. We take
$\lambda_B=\lambda_W=1$ and choose complex $\lambda_l$,
$\lambda_l=|\lambda_l|\,e^{i\,\theta_l}$ with the CP violating
parameter $\theta_l$, in order to create the EDM. Here we assume
that the couplings $|\lambda_l|$ obey the mass hierarchy of
charged leptons, $|\lambda_{\tau}|>|\lambda_{\mu}|>|\lambda_{e}|$
and we take $|\lambda_{\tau}|=1$, $|\lambda_{\mu}|=0.1$ and
$|\lambda_{e}|=0.005$.

In the first part of the calculation we restrict the CP violating
parameter $\theta_\mu$ by assuming that the antisymmetric
unparticle tensor contribution to muon anomalous magnetic moment
reaches to the experimental upper limit  $a_{\mu}=10^{-9}$ and we
study its contribution to the EDM of muon $d_\mu$. Furthermore we
predict the EDMs of electron and tau lepton and estimate the
acceptable values of the free parameters by taking the
intermediate numerical value of the CP violating parameter, namely
$sin\theta_e=sin\theta_\tau=0.5$. Finally we study the  CP
violating parameter dependence of EDMs.

\begin{figure}[htb]
\vskip -3.0truein \centering \epsfxsize=6.8in
\leavevmode\epsffile{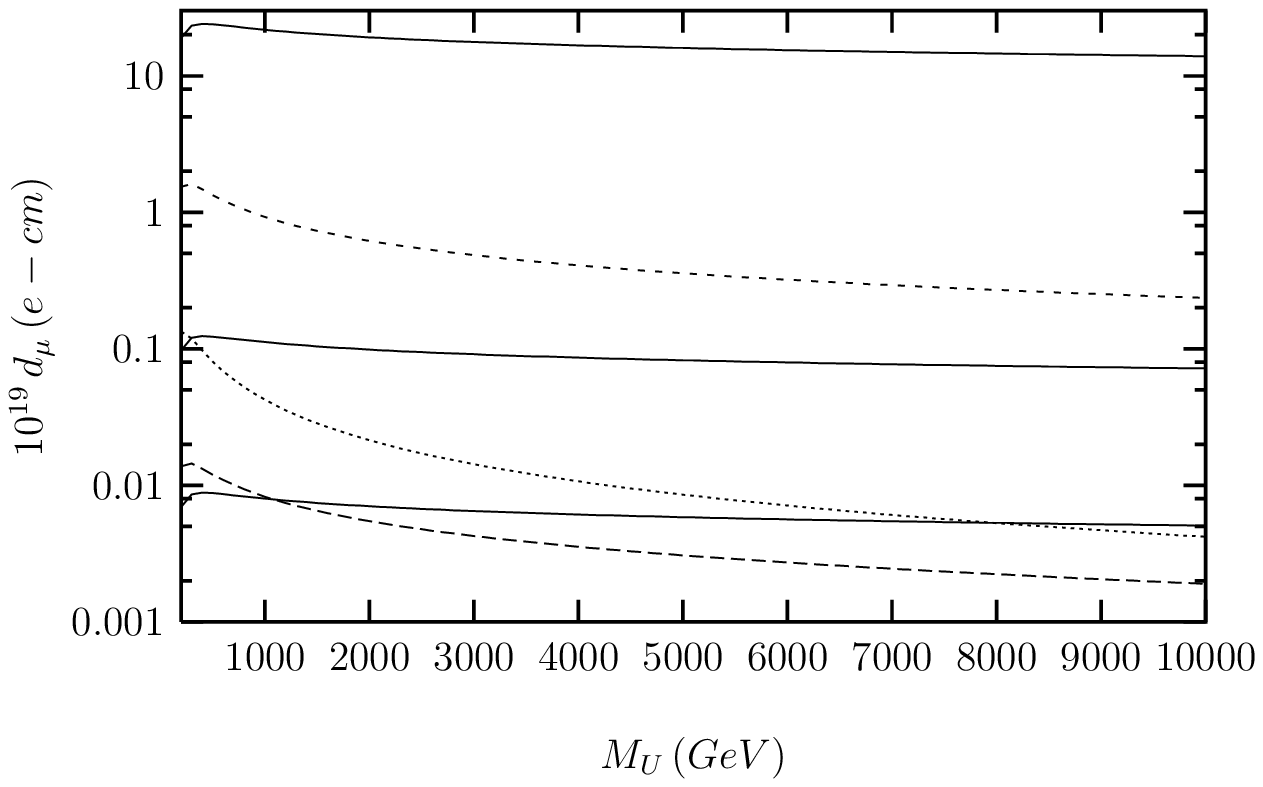} \vskip -3.0truein
\caption[]{$d_{\mu}$ with respect to $M_U$  for
$a^{U}_{\mu}=10^{-9}$. Upper-lower-the lowest solid (dashed-long
dashed; dotted) line represents the EDM for $d_U=1.1$,
$r_U=0.40-0.10-0.05$ ($d_U=1.3$, $r_U=0.40-0.10$; $d_U=1.5$,
$r_U=0.40$).} \label{muEDMMu}
\end{figure}

In  Fig.\ref{muEDMMu}, we present $M_U$ dependence of the EDM
$d_{\mu}$ for $a^{U}_{\mu}=10^{-9}$ and different values of the
scale parameter $d_U$ and the ratio $r_U=\frac{\Lambda_U}{M_U}$.
Here upper-lower-the lowest solid (dashed-long dashed; dotted)
line represents the EDM for $d_U=1.1$, $r_U=0.40-0.10-0.05$
($d_U=1.3$, $r_U=0.40-0.10$; $d_U=1.5$, $r_U=0.40$). It is
observed that $d_{\mu}$ is strongly sensitive to the ratio $r_U$
and the increasing values of $r_U$ causes the enhancement in
$d_{\mu}$. To reach the current experimental limit $r_U$ must be
at least of the order of $r_U\sim 10^{-1}$ if the scaling
dimension satisfies $d_U> 1.1$. For larger values of $d_U$ the
higher values of $r_U$ are accepted. The dependence of $d_\mu$ to
the mass scale $M_U$ is also strong especially for the large
values of the scaling dimension and it decreases more than one
order in the range $10^3\, GeV < M_U < 10^4\, GeV$ for $d_U \sim
1.5$ and more.

\begin{figure}[htb]
\vskip -3.0truein \centering \epsfxsize=6.8in
\leavevmode\epsffile{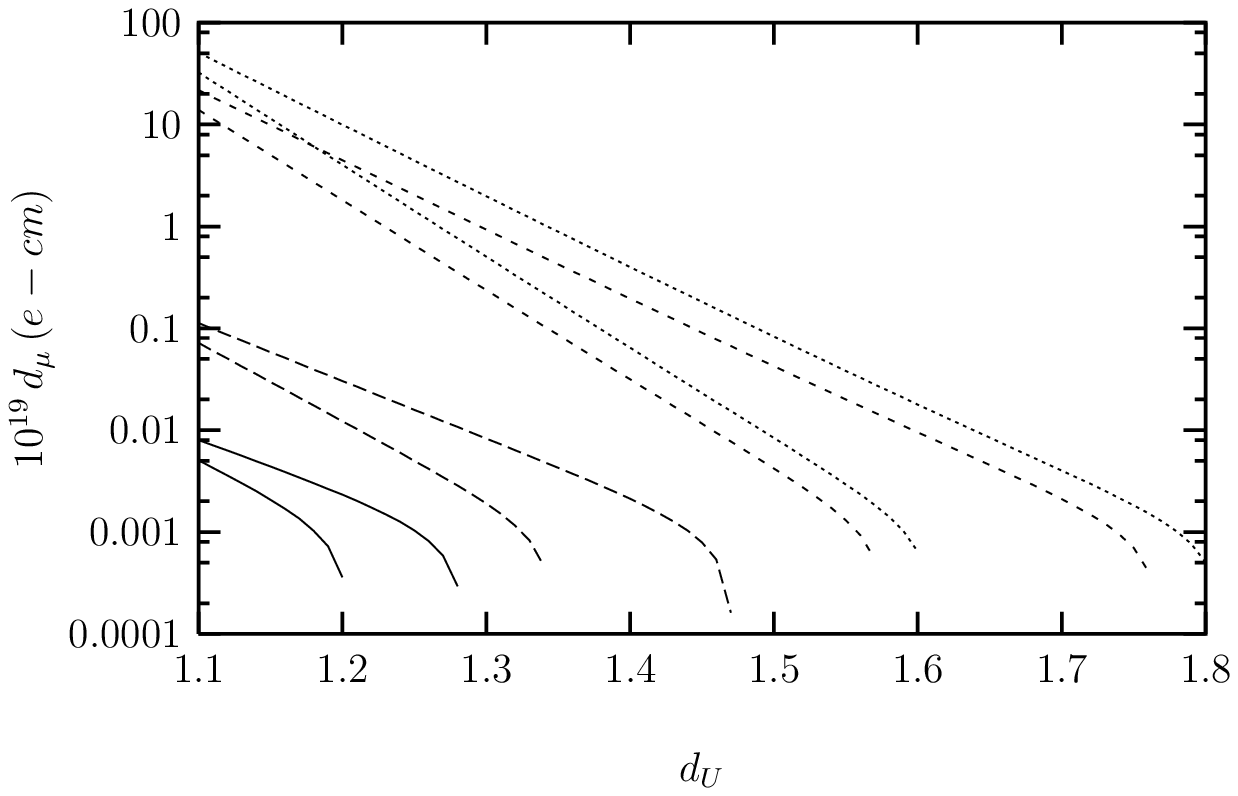} \vskip -3.0truein
\caption[]{$d_{\mu}$ with respect to the scale parameter $d_U$ for
$a^{U}_{\mu}=10^{-9}$. Here upper-lower solid (long dashed;
dashed; dotted) line represents the EDM for
$r_U=0.05$,\,$M_U=10^3\,GeV$-$r_U=0.05$,\,$M_U=10^4\,GeV$
($r_U=0.1$,\,$M_U=10^3\,GeV$-$r_U=0.1$,\,$M_U=10^4\,GeV$;
$r_U=0.4$,\,$M_U=10^3\,GeV$-$r_U=0.4$,\,$M_U=10^4\,GeV$;
$r_U=0.5$,\,$M_U=10^3\,GeV$-$r_U=0.5$,\,$M_U=10^4\,GeV$).}
\label{muEDMdu}
\end{figure}

\begin{figure}[htb]
\vskip -3.0truein \centering \epsfxsize=6.8in
\leavevmode\epsffile{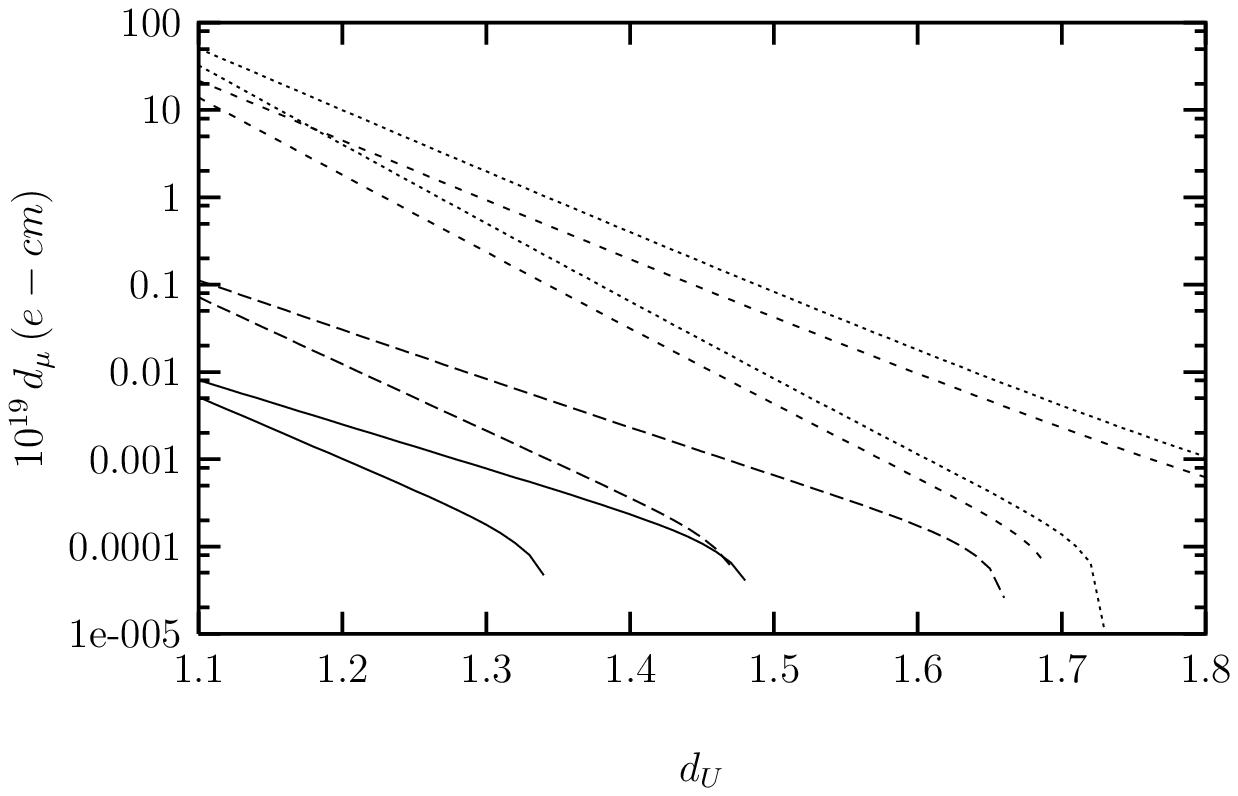} \vskip -3.0truein \caption[]{The
same as Fig. \ref{muEDMdu} but for $a^{U}_{\mu}=10^{-10}$.}
\label{muEDMdu2}
\end{figure}

Fig.\ref{muEDMdu} and Fig.\ref{muEDMdu2} are devoted to $d_{\mu}$
with respect to the scale parameter $d_U$ for
$a^{U}_{\mu}=10^{-9}$ and $a^{U}_{\mu}=10^{-10}$, respectively.
Here upper-lower solid (long dashed; dashed; dotted) line
represents the EDM for
$r_U=0.05$,\,$M_U=10^3\,GeV$-$r_U=0.05$,\,$M_U=10^4\,GeV$
($r_U=0.1$,\,$M_U=10^3\,GeV$-$r_U=0.1$,\,$M_U=10^4\,GeV$;
$r_U=0.4$,\,$M_U=10^3\,GeV$-$r_U=0.4$,\,$M_U=10^4\,GeV$;
$r_U=0.5$,\,$M_U=10^3\,GeV$-$r_U=0.5$,\,$M_U=10^4\,GeV$). For the
decreasing values of the ratio $r_U$ $d_U$ becomes more restricted
and with its the increasing values the current experimental value
can be reached. If the contribution of the antisymmetric tensor
unparticle to the anomalous magnetic moment of muon  is taken as
$a^{U}_{\mu}=10^{-10}$ (see Fig.\ref{muEDMdu2}) the restriction of
$d_U$ is more relaxed and for higher values of the ratio $r_U$ it
would be possible to reach the current experimental value of
$d_{\mu}$ similar to the previous case.

\begin{figure}[htb]
\vskip -3.0truein \centering \epsfxsize=6.8in
\leavevmode\epsffile{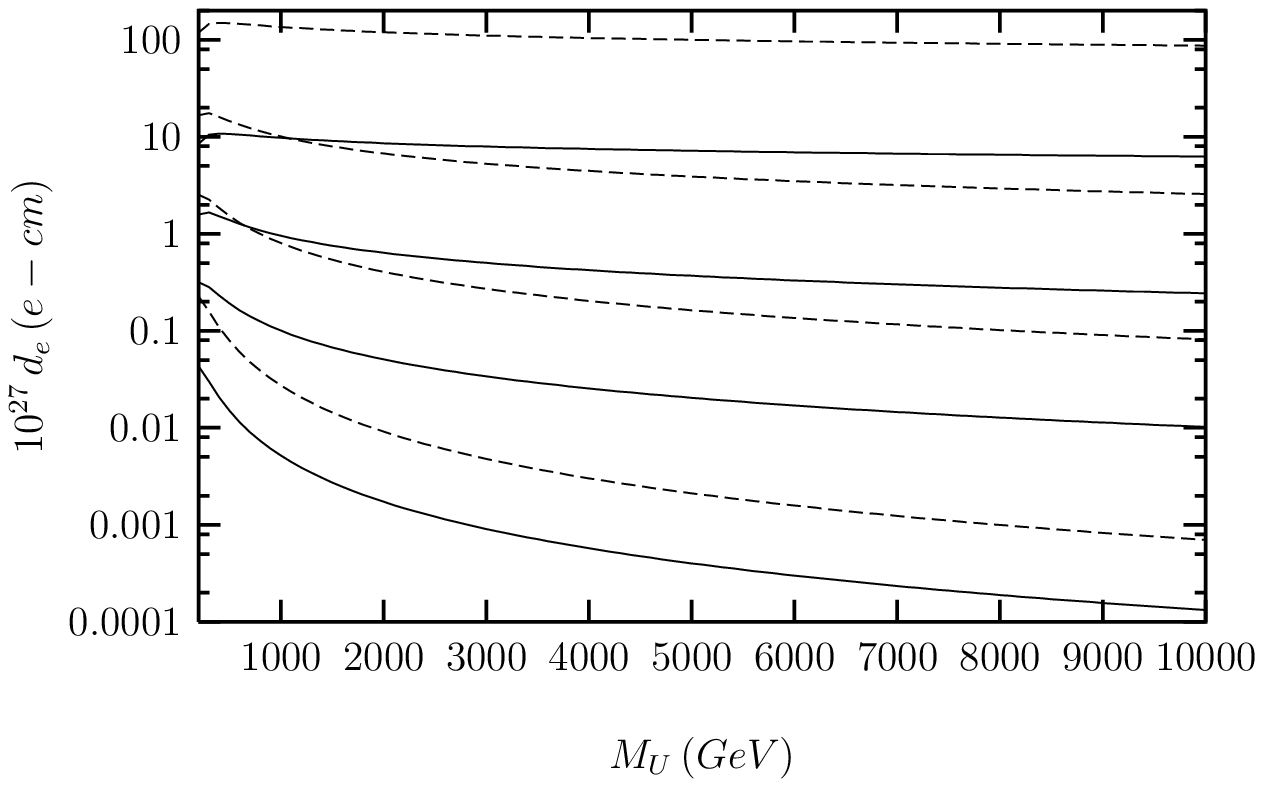} \vskip -3.0truein \caption[]{$d_e$
with respect to $M_U$ for $sin\theta_e=0.5$. Here the upper
most-upper-lower-the lowest solid; dashed line represents $d_e$
for $d_U=1.1-1.3-1.5-1.8$, $r_U=0.05$; $r_U=0.10$.} \label{eEDMMu}
\end{figure}

\begin{figure}[htb]
\vskip -3.0truein \centering \epsfxsize=6.8in
\leavevmode\epsffile{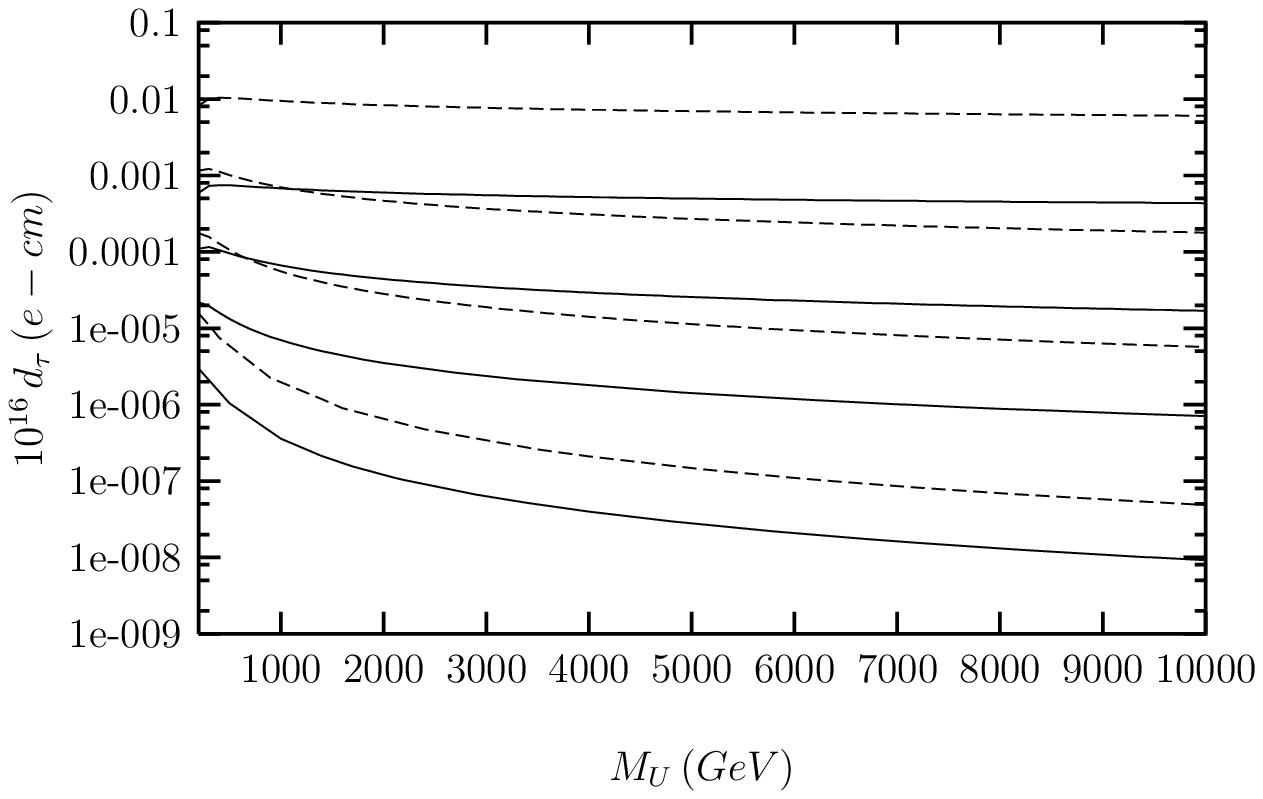} \vskip -3.0truein \caption[]{The
same as Fig. \ref{eEDMMu} but for $d_{\tau}$ and
$sin\theta_\tau=0.5$.} \label{tauEDMMu}
\end{figure}

\begin{figure}[htb]
\vskip -3.0truein \centering \epsfxsize=6.8in
\leavevmode\epsffile{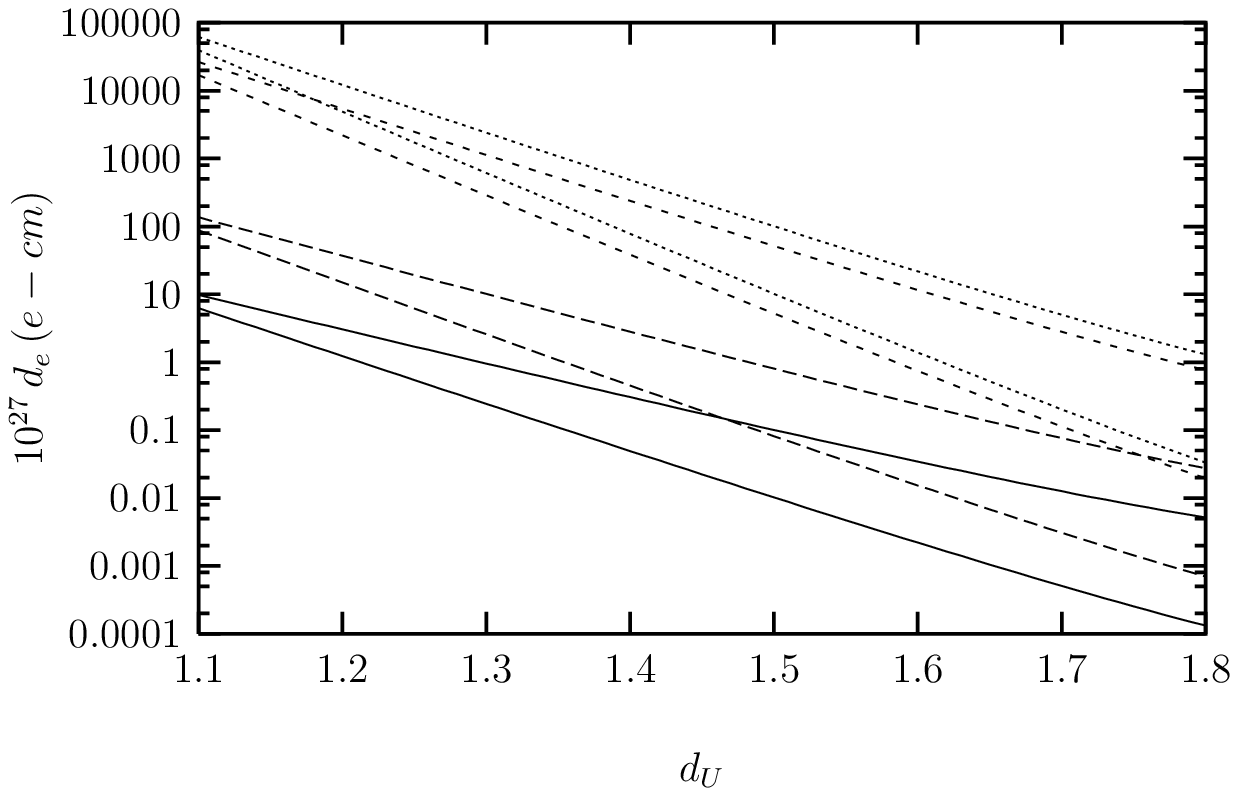} \vskip -3.0truein \caption[]{$d_e$
with respect to the scale parameter $d_U$. Here upper-lower solid
(long dashed; dashed; dotted) line represents $d_e$ for
$r_U=0.05$,\,$M_U=10^3\,GeV$-$r_U=0.05$,\,$M_U=10^4\,GeV$
($r_U=0.1$,\,$M_U=10^3\,GeV$-$r_U=0.1$,\,$M_U=10^4\,GeV$;
$r_U=0.4$,\,$M_U=10^3\,GeV$-$r_U=0.4$,\,$M_U=10^4\,GeV$;
$r_U=0.5$,\,$M_U=10^3\,GeV$-$r_U=0.5$,\,$M_U=10^4\,GeV$).}
\label{eEDMdu}
\end{figure}

\begin{figure}[htb]
\vskip -3.0truein \centering \epsfxsize=6.8in
\leavevmode\epsffile{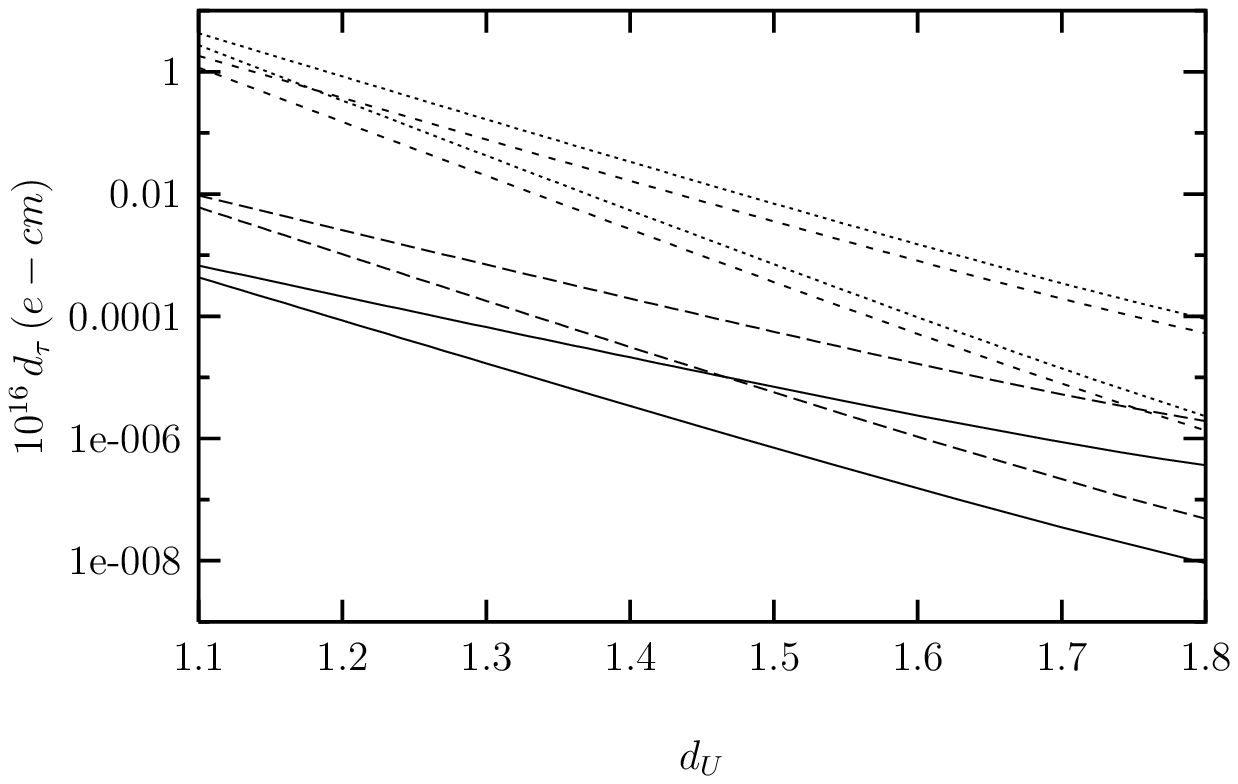} \vskip -3.0truein \caption[]{The
same as the Fig.\ref{eEDMdu} but for $d_\tau$.} \label{tauEDMdu}
\end{figure}

Fig.\ref{eEDMMu} (\ref{tauEDMMu}) represents $M_U$ dependence of
the EDM $d_e$ ($d_\tau$) for $sin\theta_e=0.5$
($sin\theta_\tau=0.5$) and for different values of the scale
parameter $d_U$ and the ratio $r_U$. Here the upper
most-upper-lower-the lowest solid; dashed line represents the
$d_e$ ($d_\tau$) for $d_U=1.1-1.3-1.5-1.8$, $r_U=0.05$;
$r_U=0.10$. We see that the increasing values of $M_U$ ($r_U$)
cause the decrease (increase) in the EDM. The current experimental
limit of $d_e$ is reached for $r_U$ which is at the order of the
magnitude of $10^{-2}$ in the case of small values of the scaling
dimension $d_U$. $r_U$ can take the values of the order of
$10^{-1}$ for $1.3< d_U < 1.5$. This can be seen also in
Fig.\ref{eEDMdu} which represents $d_U$ dependence of $d_e$ where
upper-lower solid (long dashed; dashed; dotted) line represents
the EDM for
$r_U=0.05$,\,$M_U=10^3\,GeV$-$r_U=0.05$,\,$M_U=10^4\,GeV$
($r_U=0.1$,\,$M_U=10^3\,GeV$-$r_U=0.1$,\,$M_U=10^4\,GeV$;
$r_U=0.4$,\,$M_U=10^3\,GeV$-$r_U=0.4$,\,$M_U=10^4\,GeV$;
$r_U=0.5$,\,$M_U=10^3\,GeV$-$r_U=0.5$,\,$M_U=10^4\,GeV$). For the
large values of the ratio $r_U$ the scaling dimension $d_U$ must
be near $d_U\sim 2.0$ in order to get the current experimental
value of $d_e$. On the other hand Fig.\ref{tauEDMMu} shows that
one needs the ratio $r_U\sim 0.5$ and the small values of the
scaling dimension, $d_U \sim 1.1$ in order to reach the current
experimental value of $d_\tau$ (see also Fig.\ref{tauEDMdu} which
is the same as the Fig.\ref{eEDMdu} but for $d_\tau$).

\begin{figure}[htb]
\vskip -3.0truein \centering \epsfxsize=6.8in
\leavevmode\epsffile{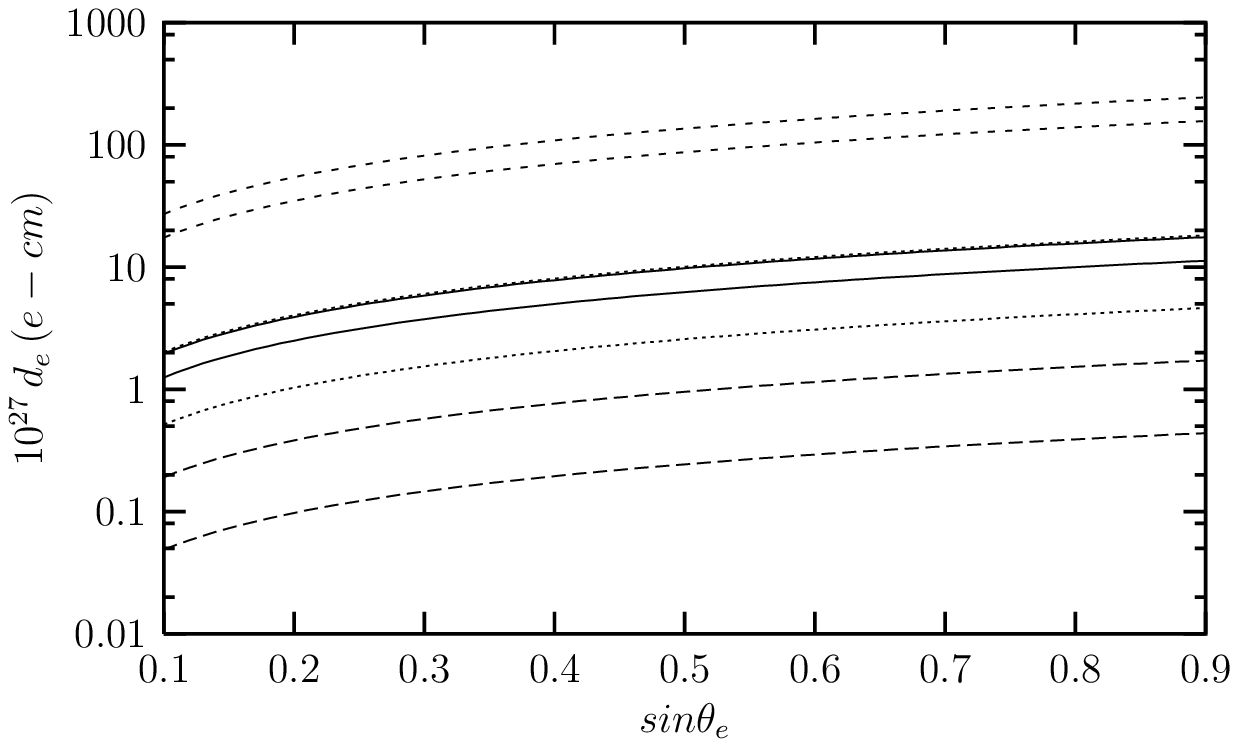} \vskip -3.0truein \caption[]{
$d_e$ with respect to $sin\theta_e$. Here upper-lower solid; long
dashed; dashed; dotted line represents  $d_e$  for
$M_U=10^3\,GeV$-$M_U=10^4\,GeV$, $r_U=0.05$, $d_U=1.1$;
$r_U=0.05$, $d_U=1.3$; $r_U=0.1$, $d_U=1.1$; $r_U=0.1$, $d_U=1.3$.
.} \label{eEDMsin}
\end{figure}

\begin{figure}[htb]
\vskip -3.0truein \centering \epsfxsize=6.8in
\leavevmode\epsffile{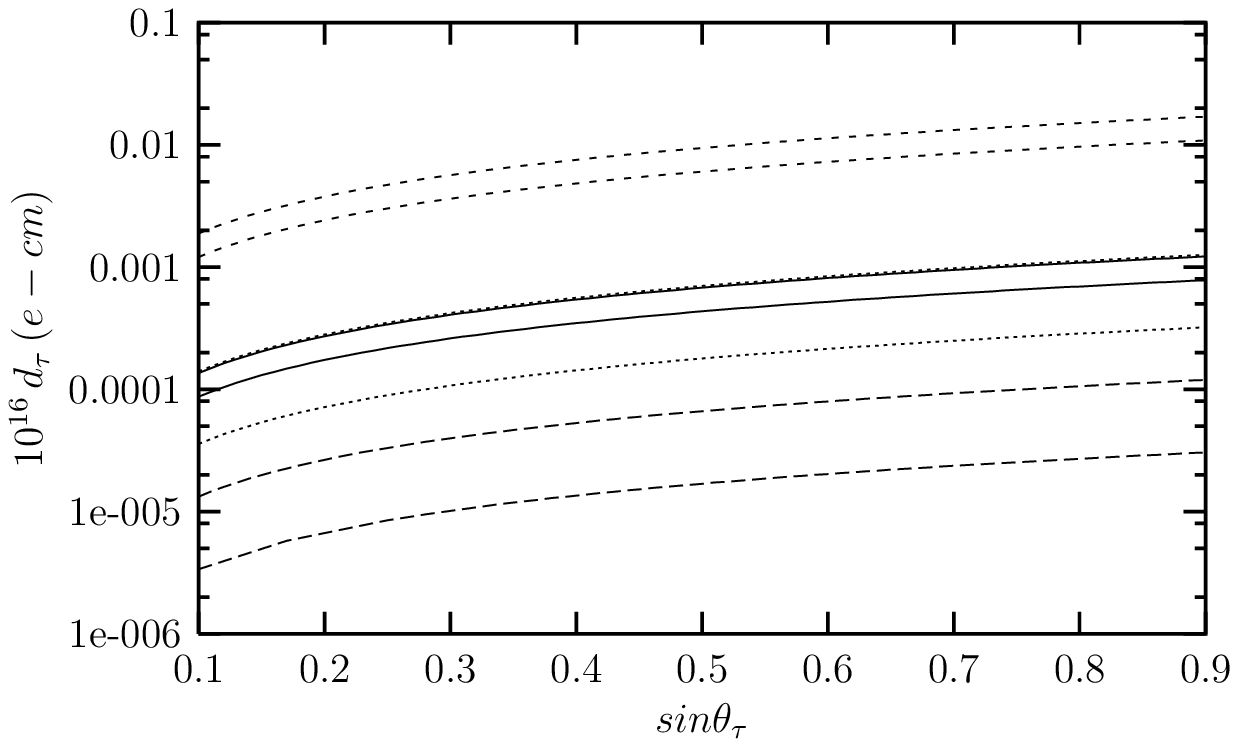} \vskip -3.0truein
\caption[]{The same as Fig. \ref{eEDMsin} but for $d_{\tau}$ and
with respect to $sin\theta_\tau$.} \label{tauEDMsin}
\end{figure}

Finally we plot the EDM $d_e$ ($d_\tau$) with respect to the CP
violating parameter $sin\theta_e$ ($sin\theta_\tau$) in
Fig.\ref{eEDMsin} (\ref{tauEDMsin}). For both figures upper-lower
solid; long dashed; dashed; dotted line represents\footnote{Notice
that the dotted line which represents $r_U=0.1$, $M_U=10^3\,GeV$,
$d_U=1.3$ almost coincides with the one which represents
$r_U=0.05$, $M_U=10^3\,GeV$, $d_U=1.1$ and it is not observed in
the figure} the $d_e$ ($d_\tau$) for
$M_U=10^3\,GeV$-$M_U=10^4\,GeV$, $r_U=0.05$, $d_U=1.1$;
$r_U=0.05$, $d_U=1.3$; $r_U=0.1$, $d_U=1.1$; $r_U=0.1$, $d_U=1.3$.
These figures show that $d_e$ and $d_\tau$ are enhanced at least
one order in the range of the CP violating parameter, $0.1 <
sin\theta_\tau < 0.9$

Now we would like to summarize our results: The charged lepton
EDMs are strongly sensitive to the parameters used, namely the
scaling dimension $d_U$, the ratio $r_U$ and the mass scale $M_U$.
We observe that the experimental current limits of $d_e$ and
$d_\mu$ are reached in the case that the ratio $r_U$ lies in the
range of $0.05-0.20$ and the scaling dimension $d_U$ is near
$1.1-1.2$. However for the current experimental value of $d_\tau$
the ratio must reach to the values $r_U\sim 0.5$ for the small
values of the scaling dimension, $d_U \sim 1.1$.

For completeness, we compare the theoretical framework and the
numerical results of the present work with the study
\cite{IltanUnp} which is related to the contribution of scalar
unparticle on the charged lepton EDM. In the present case the
tensor unparticle contribution is in the tree level, however in
\cite{IltanUnp} the scalar unparticle contribution is at one loop
level. In addition to this, in the present work, we assume that
the scale invariance is broken at some scale $\mu$ after the
electroweak symmetry breaking and, therefore,  the antisymmetric
tensor unparticle propagator is modified. In \cite{IltanUnp} the
scale invariance is intact and the propagator is the original one.
In both cases the charged lepton EDMs are strongly sensitive to
the scaling dimension $d_U$ and the experimental current limit of
$d_e$ can be reached in the range $1.6\leq d_U \leq 1.8$ (near
$1.1-1.2$) for scalar unparticle mediation (tensor unparticle
mediation). For $d_\mu$ and $d_\tau$ the current limits are
reached for the small values of the scale $d_U$, $d_U \leq 1.1$,
for both cases.

Hopefully, with in future more accurate measurements of the lepton
EDMs it would be possible to eliminate this discrepancy. These new
measurements will give strong information about the role of
unparticle scenario on the CP violation mechanism and the nature
of unparticles.
\newpage
{\Large \textbf{Appendix}}
\\ \\
Here we would like to present the calculation of the charged
lepton EDM (see eq.(\ref{EDMtensor})) by using the effective
lagrangian given in eq.(\ref{lagrangiantensor}). The first
(second) term in the effective lagrangian drives the $O^{\mu\nu}_U
\rightarrow A_\nu$ transition which is carried by the vertex
\begin{eqnarray}
2\,i\,\frac{g'\,c_W\,\lambda_B}{\Lambda_2^{d_U-2}}\,k_\mu\,\epsilon_\nu
O^{\mu\nu}_U\,\,
(-i\,\frac{g\,v^2\,s_W\,\lambda_W}{2\,\Lambda_4^{d_U}}\,k_\mu\,\epsilon_\nu
O^{\mu\nu}_U)\, , \nonumber
\end{eqnarray}
where $\epsilon_\nu$ is the outgoing photon four polarization
vector. On the other hand the third term in the effective
lagrangian results in the vertex
\begin{eqnarray}
\frac{y_l\,v}{\sqrt{2}\,\Lambda_4^{d_U}}\,(\lambda_l-\lambda_l^*)\,
\bar{l}\,\gamma_5\,\sigma_{\mu\nu}\,l\, , \nonumber
\end{eqnarray}
which creates the EDM interaction. Finally these two vertices are
connected by the tensor unparticle propagator (see
eq.(\ref{propagatormu})) and, by extracting the coefficient of
$i\,\bar{l}\,\gamma_5 \,\sigma^{\mu\nu}\,l\, F_{\mu\nu}$, one gets
the EDM of charged leptons as in eq.(\ref{EDMtensor}). Now we give
a brief explanation how to obtain the tensor unparticle
propagator. The starting point is the scalar unparticle propagator
which is obtained by respecting the scale invariance. The two
point function of scalar unparticle operators reads
\begin{eqnarray}
<0|\Big(O_U(x)\,O_U(0)\Big)0>= \int\,\frac{d^4P}{(2\,\pi)^4}\,
e^{-iP.x}\,\rho(P^2)
 \, ,
\label{twopointfunct}
\end{eqnarray}
where $\rho(P^2)$ is the spectral density:
\begin{eqnarray}
\rho(P^2)= A_{d_U}\,\theta(P^0)\,\theta(P^2)\,(P^2)^\xi \, .
\label{spectraldensty}
\end{eqnarray}
The scale invariance\footnote{The spectral density is invariant
under the scale transformation $x\rightarrow s\,x$ and
$O_U(s\,x)\rightarrow s^{-d_U}\,O_U(x)$.} requires a restriction
on the parameter $\xi$, $\xi=d_U-2$, and, therefore, $\rho(P^2)$
becomes
\begin{eqnarray}
\rho(P^2)= A_{d_U}\,\theta(P^0)\,\theta(P^2)\,(P^2)^{d_U-2} \, .
\label{spectraldensty2}
\end{eqnarray}
Here the factor $A_{d_U}$ reads
\begin{eqnarray}
A_{d_U}=\frac{16\,\pi^{5/2}}{(2\,\pi)^{2\,d_U}}\,
\frac{\Gamma(d_U+\frac{1}{2})} {\Gamma(d_U-1)\,\Gamma(2\,d_U)} \,
, \nonumber
\end{eqnarray}
in order to get the phase space of $d_U$ massless particles, i.e.,
unparticle stuff having the scale dimension $d_U$ can be
represented as non-integral number $d_U$ of invisible particles
\cite{Georgi1, Georgi2, Cheung2}. Finally, by using spectral
formula, the scalar unparticle propagator is obtained as
\cite{Georgi2, Cheung2}
\begin{eqnarray}
\int\,d^4x\,
e^{iP.x}<0|T\Big(O_U(x)\,O_U(0)\Big)0>=i\frac{A_{d_U}}{2\,\pi}\,
\int_0^{\infty}\!\!\!
ds\,\frac{s^{d_U-2}}{P^2-s+i\epsilon}\!=i\,\frac{A_{d_U}}
{2\,sin\,(d_U\pi)}\,(-P^2-i\epsilon)^{d_U-2}. \label{propagators}
\end{eqnarray}
Notice that for $P^2>0$, the function
$\frac{1}{(-P^2-i\epsilon)^{2-d_U}}$ in eq. (\ref{propagators})
reads
\begin{eqnarray}
\frac{1}{(-P^2-i\epsilon)^{2-d_U}}\rightarrow
\frac{e^{-i\,d_U\,\pi}}{(P^2)^{2-d_U}} \, , \label{strongphase}
\end{eqnarray}
which shows that there exists a non-trivial phase due to the
non-integral scaling dimension. In the case of tensor unparticle
one needs a projection operator $\Pi_{\mu\nu\alpha\beta}=
\frac{1}{2}(g_{\mu\alpha}\,g_{\nu\beta}-g_{\nu\alpha}\,g_{\mu\beta})$
which contains the transverse and longitudinal parts and one gets
the propagator of antisymmetric tensor unparticle as
\begin{eqnarray}
\int\,d^4x\,
e^{ipx}\,<0|T\Big(O^{\mu\nu}_U(x)\,O^{\alpha\beta}_U(0)\Big)0>=
i\,\frac{A_{d_U}}
{2\,sin\,(d_U\pi)}\,\Pi^{\mu\nu\alpha\beta}(-p^2-i\epsilon)^{d_U-2}
\, . \nonumber
\end{eqnarray}
\newpage
\newpage

\begin{thebibliography}{1}
%
\bibitem{Commins}
B. C. Regan, E. D. Commins, C. J. Schmidt, and D. DeMille, {\it
Phys. Rev. Lett. } {\bf  88}, 071805 (2002).
%
\bibitem{Bailey} J. Bailey, et al, {\it Journ. Phys.} {\bf G4},
345 (1978).
%
\bibitem{KInami}  Belle collaboration, K.Inami, et al,
{\it Phys. Lett.} {\bf B551}, 16 (2003).
%
\bibitem{Schmidt} C. R. Schmidt and M. E. Peskin, {\it Phys. Rev. Lett.}
{\bf 69} (1992) 410.
%
\bibitem{Iltmuegam} E. Iltan, {\it Phys. Rev.} {\bf D64}, 013013 (2001).
%
\bibitem{Bhaskar} B. Dutta, R. N. Mohapatra, {\it Phys. Rev.}
{\bf D68}, 113008 (2003).
%
\bibitem{IltanExtrEDM} E. Iltan, {\it JHEP} {\bf 0404}, 018 (2004).
%
\bibitem{IltanSplitEDM} E. Iltan, {\it Eur. Phys.} {\bf C44 }, 411 (2005).
%
\bibitem{IltanNonCom} E. Iltan, {\it JHEP} {\bf 065}, 0305 (2003).
%
\bibitem{Georgi1}H. Georgi, {\it Phys. Rev.  Lett.}  {\bf 98}, 221601 (2007).
%
\bibitem{Georgi2} H. Georgi, {\it Phys. Lett.} {\bf B650}, 275 (2007).
%
\bibitem{Cheung2} K. Cheung, W. Y. Keung and T. C. Yuan,
{\it Phys. Rev.}  {\bf D76}, 055003 (2007).
%
\bibitem{BankZaks} T. Banks, A. Zaks, {\it Nucl. Phys.} {\bf B196}, 189 (1982).
%
\bibitem{Rajaraman} M. Bander, J. L. Feng, A. Rajaraman, Y.
Shirman, {\it Phys. Rev.} {\bf D76}, 115002 (2007).
%
\bibitem{Tae}T. Hur, P. Ko, X. H. Wu, {\it Phys. Rev.}
 {\bf D76}, 096008 (2007)
%
\bibitem{IltanUnp} E. Iltan, {\it Int. J. Mod. Phys.} {\bf A24 }, 2729 (2009).
%
\bibitem{PJFox}  P. J. Fox, A. Rajaraman, Y. Shirman,
{\it Phys. Rev.}  {\bf D76}, 075004 (2007).
%
\bibitem{Kikuchi} T. Kikuchi, N. Okada, {\it Phys. Lett.} {\bf B661}, 360
(2008).
%
\bibitem{ADelgado}A. Delgado, J. R. Espinosa, J. M. No and M. Quiros,
{\it JHEP} {\bf 0804}, 028 (2008)
%
\bibitem{ARajaraman}  A. Rajaraman, {\it Phys. Lett.} {\bf B671}, 411
(2009).
%
%
\bibitem{Grinstein}  B. Grinstein,  K. A. Intriligator, I. Z. Rothstein
{\it Phys. Lett.} {\bf B662}, 367 (2008).
%
\end{thebibliography}
\end{document}